\documentclass[a4paper,10pt,twocolumn]{article}
\pdfoutput=1

\usepackage{graphicx}
\usepackage{grffile}
\usepackage{xcolor}

\usepackage{url}
\usepackage[pdftex,colorlinks=true,linkcolor=blue,citecolor=blue,urlcolor=blue]{hyperref}

\usepackage[expproduct=cdot]{siunitx} 
\usepackage{enumitem} 
\usepackage{epstopdf}

\usepackage{relsize} 
\usepackage{amsmath}
\usepackage{amssymb}
\usepackage{bbm} 
\usepackage[utopia,greekuppercase=italicized]{mathdesign}\edef\partial{\mathchar\number\partial\noexpand\!} 

\usepackage[T1]{fontenc}

\usepackage{ctable} 
\usepackage{tabularx}

\usepackage[numbers,compress]{natbib}

\definecolor{royalblue4}{HTML}{27408B} 
\definecolor{red4}{HTML}{8B0000} 
\definecolor{green4}{HTML}{008b00}

\usepackage{dblfloatfix}  
\usepackage{fixltx2e}  

\usepackage[font=footnotesize,labelfont=bf,labelsep=endash,margin=0pt]{caption} 
\usepackage[title,header]{appendix}

\newlength{\myleftmargin} 
\newlength{\mytopmargin} 
\newlength{\myrightmargin} 
\newlength{\mybottommargin} 
\usepackage[runin]{abstract}

\let\paragraphold\paragraph
\renewcommand*{\paragraph}[1]{\paragraphold{#1.}} 
\newcommand{\keywords}[1]{\vspace{2mm}\noindent\textbf{Keywords:} #1} 
\newcommand{\pagewidetitle}[3] 
{
    \twocolumn
        [
            \vskip-5mm
            \begin{@twocolumnfalse}%
                #1
                #2
                \vspace{5mm}
            \end{@twocolumnfalse}%
        ]
        #3
}

\newlength{\figurewidth}

\definecolor{pbuenzli}{HTML}{00b7eb} 
\definecolor{pbuenzlinew}{HTML}{008b00}
\definecolor{pbuen}{HTML}{A020F0}

\makeatletter
\long\def\tlist@if@empty@nTF #1{
\expandafter\ifx\expandafter\\\detokenize{#1}\\
\expandafter\@firstoftwo
\else
\expandafter\@secondoftwo
\fi
}

\newcommand*\pbuenzli[2][]{
{\color{pbuenzli}#2}
\tlist@if@empty@nTF{#1}{}{\footnote{\color{pbuenzli}\it #1}}
}
\makeatother

\usepackage{relsize}

\begin{document}
	\title{\bf Osteoblasts infill irregular pores under curvature and porosity controls: A hypothesis-testing analysis of cell behaviours}

	\renewcommand{\thefootnote}{\fnsymbol{footnote}}%
	\author{Mohd Almie Alias$^\text{a,b,}$\footnotemark[1], Pascal R Buenzli$^\text{a,c}$}
	
	\date{\normalsize \vspace{-2mm}$^\text{a}$School of Mathematical Sciences, Monash University, Clayton VIC 3800, Australia\\$^\text{b}$School of Mathematical Sciences, National University of Malaysia, 43600 Bangi, Selangor D. Ehsan, Malaysia\\$^\text{c}$School of Mathematical Sciences, Queensland University of Technology, Brisbane QLD 4001, Australia\\\vskip 1mm \normalsize	
	\today\vspace*{-5mm}}
	
	\pagewidetitle{
		\maketitle
	}{
      \begin{abstract}
		The geometric control of bone tissue growth plays a significant role in bone remodelling, age-related bone loss, and tissue engineering. However, how exactly geometry influences the behaviour of bone-forming cells remains elusive. Geometry modulates cell populations collectively through the evolving space available to the cells, but it may also modulate the individual behaviours of cells. To factor out the collective influence of geometry and gain access to the geometric regulation of individual cell behaviours, we develop a mathematical model of the infilling of cortical bone pores and use it with available experimental data on cortical infilling rates. Testing different possible modes of geometric controls of individual cell behaviours consistent with the experimental data, we find that efficient smoothing of irregular pores only occurs when cell secretory rate is controlled by porosity  rather than curvature. This porosity control suggests the convergence of a large scale of intercellular signalling to single bone-forming cells, consistent with that provided by the osteocyte network in response to mechanical stimulus. After validating the mathematical model with the histological record of a real cortical pore infilling, we explore the infilling of a population of randomly generated initial pore shapes. We find that amongst all the geometric regulations considered, the collective influence of curvature on cell crowding is a dominant factor for how fast cortical bone pores infill, and we suggest that the irregularity of cement lines thereby explains some of the variability in double labelling data as well as the overall speed of osteon infilling.

\keywords{Bone remodelling, tissue growth, osteoblast, tissue engineering, morphogenesis}
	\end{abstract}
}{
\protect\footnotetext[1]{Corresponding author. Email address: \texttt{mohdalmie@ukm.edu.my}}%
\renewcommand{\thefootnote}{\arabic{footnote}}%
}

\section{Introduction}
\label{Section_Introduction}
The geometric control of biological tissue growth has been evidenced in several tissue engineering constructs~\citep{Chen1997,Rolli2012,Paris2017}, but much of the cellular mechanisms that underlie this control still remain to be elucidated. In-vitro experimental studies track the evolution of the tissue interface and analyse correlations between growth rate and local curvature, but they report little quantitative information about the tissue-forming cells~\citep{Rumpler2008,Bidan2012,Bidan2013,Bidan2016,Ripamonti2009}. It remains unknown how tissue-forming cells are regulated individually by geometric clues such as curvature, whether in vivo or in vitro. 

Mathematical models can help link observed tissue-scale dynamics with unknown cellular activity~\citep{Alias2017}. To discover how geometric clues regulate cells individually, it is essential to extract from experimental growth rates the inevitable collective influence of geometry. Indeed, the crowding/spreading of the population of cells that is due to the shrinking/expanding space available to the cells affects tissue growth geometrically. Factoring out this collective effect is only possible using geometric reasoning in mathematical models. The cell-based model and hypothesis-testing analysis that we present in this paper for bone tissue growth allows us to factor out the collective influence of curvature on the population of osteoblasts (bone-forming cells), and thereby get insights into the geometric control of the individual behaviour of osteoblasts during the infilling of bone pores in vivo.

While our analysis is applied specifically to the infilling of bone pores, the procedure is applicable to investigate other systems in which geometric controls are present, including tissue growth in tissue engineering bioscaffolds ~\citep{OBrien2011}, trabecula fenestration ~\citep{Kinney1995}, the rate of closure of bone defects such as calvarial defects ~\citep{Schantz2012}, implant adhesion~\citep{Puleo1999}, wound healing~\citep{Rolli2012,Poujade2007}, and tumour growth~\citep{Lowengrub2010}. A number of studies investigate mechanical adaptation of bone but do not take into account the influence of the bone microstructure's local curvature. It is important that this geometric influence is taken care of (mathematically) so that the true effect of other influences, such as mechanics, can be estimated properly. A better understanding of the geometric control of bone growth is also important for interpreting bone tissue microstructures in bioarchaeology, such as for estimating archaeological age and activity~\citep{Buckberry2002,Maggiano2008,Mays2010}, for species identification~\citep{Martiniakova2006,Martiniakova2007}, and for understanding growth patterns in antler development~\citep{Price2005} and in plexiform bone. Our focus on the infilling of bone pores is motivated by its relevance to osteoporosis and by the availability of dynamic information on in-vivo bone formation from double labelling experiments.

Bone pores are created by osteoclasts (bone-resorbing cells) during the lifelong renewal of bone tissues. These pores are subsequently infilled by osteoblasts, which attach to the walls of the pore cavity and secrete new bone matrix~\citep{Martin1998}. During age-related bone loss and osteoporosis, it is known that bone porosity increases first as a result of such remodelling cavities not infilling completely, then as a result of increased resorption~\citep{Seeman2008}, but the detailed geometric and mechanical factors that control how bone pores infill are poorly understood. By increasing their size, pores may coalesce and become more irregular~\citep{Bell2001,Thomas2006,Lassen2017,Andreasen2017}. Because increased bone porosity leads to mechanically compromised bone and increased fracture risk~\citep{Schaffler1988,Currey1988,McCalden1993}, it is important to understand how osteoblasts respond to the local geometric features of remodelling cavities.

In-vivo labelling experiments give some insights into the infilling rate of remodelling cavities. The sequential administration of fluorochrome substances in an organism, such as tetracycline, alizarin, and calcein, leaves a series of fluorescent labels within bone. These labels record the location of past bone surfaces that were forming at the time of administration. Such experimental data suggests that the velocity of bone-forming surfaces in cortical bone, called matrix apposition rate (MAR)~\citep{Parfitt1983}, is proportional to the average radius $R$ of infilling cylindrical cavities~\citep{Metz2003,Lee1964,Manson1965}. However, it is unclear how the irregularity of infilling cavities may influence our interpretation of double labelling data. For perfectly symmetric pores (circular cross-section), the dependence upon $R$ may correspond to an influence of the curvature of the bone surface $1/R$, or an influence of porosity $\propto R^2$ indistinctively. For noncircular infilling pores, however, curvature and porosity are independent variables, so that their respective influence on osteoblasts can be differentiated.

By seeding cells of osteoblastic lineage onto bioscaffolds of different geometries, tissue engineering studies have suggested that the velocity of the tissue surface at the onset of formation is proportional to curvature where the tissue substrate is concave~\citep{Rumpler2008,Bidan2012,Bidan2013,Bidan2016,Guyot2014,Guyot2015,Guyot2016}. In a previous work, we have developed a mathematical model of tissue-forming cells to capture the systematic influence of local curvature on cell density due to the shrinking or expanding surface area near concavities or convexities of the interface~\citep{Alias2017}. Both the smoothing of highly curved regions of the interface and tissue deposition slowdown observed in the bioscaffold experiments in various pore shapes was explained by our model as a combination of (i) curvature-dependent changes in cell density; (ii) cell diffusion along the interface; and (iii) depletion of actively secreting cells.

In the present paper, we apply this mathematical model of tissue-forming cells to cortical bone pore infilling in order to extract mathematically from double labelling data the unavoidable influence of curvature on cell crowding or spreading. By taking care of this collective influence of curvature on cell density, we are able to examine how two individual cell behaviours, namely cell secretory rate (volume of bone formed per cell per unit time), and cell depletion rate (probability per unit time for the cell to become inactive, e.g. by undergoing apoptosis or anoikis) may depend upon curvature and porosity during the infilling of remodelling cavities. The underlying biological and physical processes involved in a curvature control of bone formation are likely to be fundamentally different from those involved in a porosity control of bone formation. This distinction could thus be important for understanding the evolution of age-related bone loss, and how to best counter it~\citep{Cowin1999,Daly2017}. To our knowledge, this is the first time that a mathematical model of cell population is used in combination with experimental data to gain insights into geometric influences at the level of individual cells.

Previous mathematical models of infilling bone pores have been developed \citep{Polig1990,Buenzli2011,Buenzli2014a,Petrov1989}. These models all assumed perfectly cylindrical geometries in which no distinction is possible between curvature and porosity. The models by~\cite{Polig1990} and~\cite{Buenzli2014a} included density concentration due to surface area shrinkage during pore infilling, and the generation of osteocytes by embedment of some of the bone-forming cells. The generation of osteocytes in arbitrary geometries was generalised by~\cite{Buenzli2015,Buenzli2016}. The novelty of the mathematical model itself presented here is to propose a comprehensive population model of bone-forming cells that includes osteocyte generation, and collective and individual geometric influences at the cell--tissue scale in arbitrary pore geometries.

\section{Materials and methods}

\label{Section_methods}

\subsection{Mathematical model}
\label{Section_model}
The remodelling of compact bone tissues is operated by self-contained groups of osteoclasts and osteoblasts that tunnel through old bone to replace it~\citep{Martin1998}. The osteoclasts first hollow out a cylindrical pore. The osteoblasts then attach to the walls of this cavity and infill the pore from outside in with concentric bone layers called lamellae, leaving a residual channel (Haversian canal) that contains vasculature, lymphatics, and nerves~\citep{Martin1998,Parfitt1994,Cowin2001,Burr2014}. The new bone structure thus formed is called an osteon.

\begin{figure}[t]
	\captionsetup{justification=centering}
	
	\centerline{
		\includegraphics[trim={0 0 0 0} ,width=0.95\linewidth,clip]{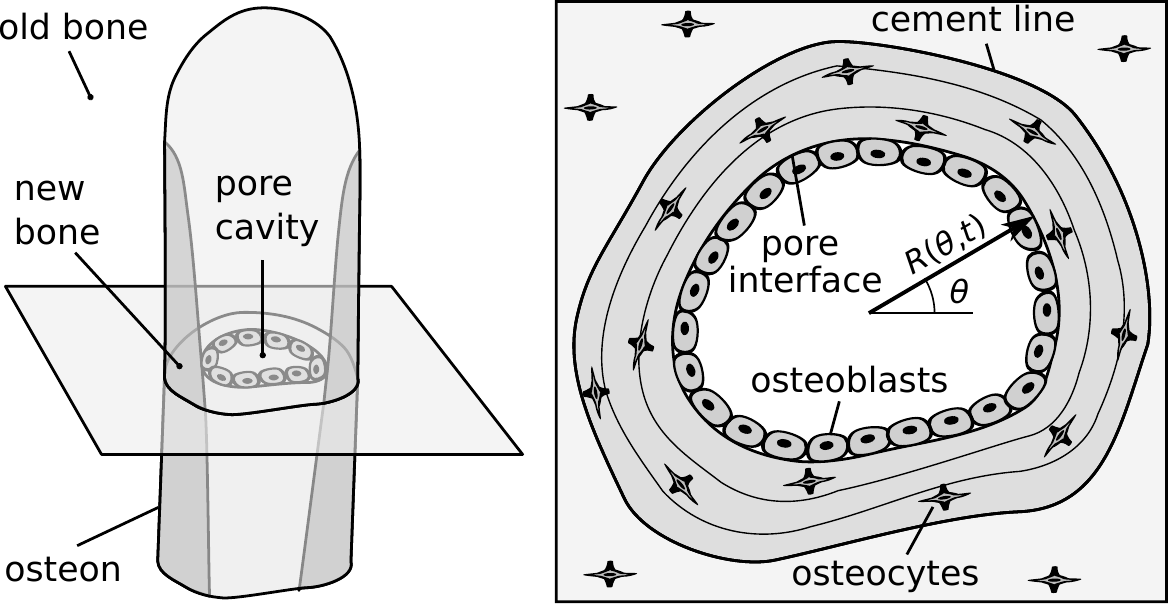}}
	
	\caption{Schematic representation of a remodelling osteon (left), and its transverse cross-section (right) showing the
		cement line (boundary between old and new bone), the pore interface described mathematically in polar coordinates by an explicit parameterisation $R(\theta,t)$, and osteoblasts described mathematically by their density $\rho$.}
	\label{fig0}
\end{figure}

Osteons form elongated cylindrical structures, so that most of the geometric regulation of pore infilling can be assumed to arise from the remodelling cavity's cross-sectional shape. We thus consider the infilling of a remodelling cavity viewed in a transverse cross section, with a pore interface $S(t)$ described by an explicit parameterisation $\theta \mapsto R(\theta, t)$ in polar coordinates (Fig. \ref{fig0}). The interface evolves by the secretion of new bone matrix by osteoblasts (bone-forming cells) lining the interface with a surface density $\rho$ (number of cells per unit surface). The normal velocity of the interface is
\begin{align}
	v=k_\text{f} \ \rho   \label{eqn_v},
\end{align}
where $k_\text{f}$ is the cell secretory rate (volume of new bone matrix secreted per cell per unit time)~\citep{Buenzli2015,Alias2017}. In the biology literature, the normal velocity of the bone interface is referred to as `matrix apposition rate' (MAR)~\citep{Parfitt1983}. Both terminologies are used interchangeably in this paper. In~\cite{Alias2017}, we developed evolution equations for the crowding and spreading of cells induced by changes in the local surface area at concavities and convexities of the interface during its evolution. Here, we extend these equations by including explicitly the formation of osteocytes by embedment of some of the osteoblasts into the bone matrix~\citep{Buenzli2015,Buenzli2016}. The evolution equations governing the pore interface radius $R(\theta,t)$ and surface density $\rho(\theta,t)$ of osteoblasts are~\citep{Alias2017}: 
\begin{align}
	R_t &= -v\sqrt{1+\left(\frac{R_\theta}{R}\right)^2} \label{eqn_R} \\
	\rho_t &= -\rho v \kappa - \frac{\rho_\theta}{R} v \cos \alpha + D\left(\frac{\rho_{\theta \theta}}{g^2} - \frac{\rho_\theta}{R} \left[\frac{2}{g}-\kappa\right] \cos\alpha \right) \nonumber\\
	& \quad - \text{Ot}_\text{f}\, v - A\rho   \label{eqn_Rho},
\end{align}
where $g=R\sqrt{1+\left(R_\theta/R\right)^2}$ is the metric, or local stretch of the parameterisation; $\cos \alpha = \boldsymbol{n} \cdot \boldsymbol{\hat{\theta}}=R_\theta/g$ is the projection of the unit normal vector $\boldsymbol{n}$ of the interface onto $\boldsymbol{\hat{\theta}}=(-\sin \theta,\cos \theta)$; and

\begin{align}
	\kappa=-\frac{R^2-RR_{\theta \theta}+2R_\theta^2}{g^3}
\end{align}
is the local curvature in polar coordinates. Curvature is taken to be negative on concave portions of the bone substrate, and positive on convex portions of the bone substrate.

The first term in the right-hand side of Eq.~(\ref{eqn_Rho}) describes the systematic dilution or concentration of osteoblasts induced by the curvature of the interface. The second term represents the transport of osteoblasts perpendicularly to the interface but measured with respect to the coordinate $\theta$. The third term is proportional to the Laplace--Beltrami operator expressed in polar coordinates. This term describes the diffusion of osteoblasts parallel to the interface with constant diffusivity $D$. This surface diffusion acts toward homogenising osteoblast density and may be interpreted as osteoblast migration along the bone surface in the direction opposite to the local density maximum. More biologically, it may also be understood as a reorganisation of the cells on the bone surface to maintain reasonable densities, arising from spring-like cell--cell contact interactions~\citep{Meineke2001,Murray2009}. The sink term $-\text{Ot}_\text{f}\, v$ represents the decrease in cell density due to the embedment of osteoblasts in bone matrix, where $\text{Ot}_\text{f} = \text{Ot}_\text{f}\big(R(\theta,t), \theta\big)$ is the instantaneous density of osteocytes (number of cells per unit volume) generated at the moving bone formation front~\citep{Buenzli2015}. The radial dependence of $\text{Ot}_\text{f}$ denotes a possible spatial dependence of osteocyte density within the osteon, which could be provided from experimental measurements. It is known that the generation of this density does not depend explicitly on the curvature of the interface ~\citep{Buenzli2015}. Here we assume for simplicity that osteocyte density is constant, $\text{Ot}_\text{f} \approx 31,250\,\text{mm}^{-3}$  ~\citep{Parfitt1983,Buenzli2014a,Buenzli2015,Hannah2010,Graham2013,FranzOdendaal2006,Dallas2010}. Finally, the sink term $-A \rho$ in Eq.~\eqref{eqn_Rho} represents depletion from the pool of active osteoblasts other than by differentiation into osteocytes, occurring at rate $A$ (in day$^{-1}$). This depletion may represent cell death (such as apoptosis), or detachment from the bone surface. Bone surface area shrinks during bone formation, which tends to increase osteoblast density, and thereby also tends to increase interface velocity. However, bone deposition is observed to slow down during formation. This means that many cells are removed from the pool of active osteoblasts during bone formation by such a depletion mechanism~\citep{Parfitt1994,Buenzli2014a}.

\paragraph{Individual cell behaviours} Individual cell behaviours are represented in Eqs~\ref{eqn_v}--\ref{eqn_Rho} by the cell secretory rate $k_\text{f}$, cell depletion rate $A$, and cell diffusivity $D$. We will assume that secretory rate $k_\text{f}$ and cell depletion rate $A$ may depend on the local geometry of the interface, but will assume constant diffusivity $D$. Cell secretory rate is expected to scale with cell body volume~\citep{Zallone1977}, which is likely to depend on the local curvature of the bone substrate, e.g., via cell density. Geometry may also control the reduction in cell secretory rate when osteoblasts become living, quiescent cells lining the bone surface at the end of bone formation, when the remaining pore is about 40\,$\muup$m in diameter~\citep{Parfitt1994}. Similarly, curvature and its effect on cell density may influence osteoblast apoptosis or detachment from the bone surface. In contrast, osteoblast diffusion parallel to the bone surface is expected to be small and only weakly dependent on curvature. Indeed, active osteoblasts form a confluent layer of cells on the bone surface~\citep{Parfitt1994}. Their cellular protrusions link with bone-matrix-embedded osteocytes, but osteocytes density is generated independently of an explicit geometric regulation~\citep{Buenzli2015}.

To gain insights into the geometric regulation of the individual cell behaviours $k_\text{f}$ and $A$, we use the mathematical model in Eqs~\eqref{eqn_v}--\eqref{eqn_Rho} in two steps:
\begin{enumerate}
	\item \textbf{Circular pore geometry.} We first consider perfectly circular infilling remodelling cavities. In this case, cell diffusion is irrelevant, and both curvature and porosity are related to the infilling pore radius $R(t)$. Direct experimental data is available from the published literature on $v(R)$, $\rho(R)$, and $k_\text{f}(R)$~\citep{Marotti1976}, see Section~\ref{Section_data}. The data $k_\text{f}(R)$ is used as input to the model in Eqs~\eqref{eqn_v} and~\eqref{eqn_Rho}, and we determine how cell depletion rate $A(R)$ must depend on $R$ for the model to match the experimental data on $v(R)$ and $\rho(R)$.
	
	\item \textbf{Noncircular pore geometry.} Knowing the dependence upon $R$ of $k_\text{f}(R)$ and $A(R)$ in circular geometry, we then seek to extrapolate this geometric regulation in noncircular geometries, where curvature $\kappa$ and porosity $\phi$ are independent variables. To this end, we propose four models of geometric regulation of $k_\text{f}$ and $A$ that are consistent with their expression in circular geometry, and explore the infilling behaviour of irregular pores to test these models. The four models considered are:
	
	\begin{description}
		
		\item[Model 1:] $k_\text{f}(\kappa)$ and $A(\kappa)$. Both the secretory behaviour of an osteoblast and its likelihood to become nonactive depend on the local curvature of the bone surface that it occupies.
		
		\item[Model 2:] $k_\text{f}(\kappa)$ and $A(\phi)$. The secretory behaviour of an osteoblast depends on the local curvature of the bone surface that it occupies, but its likelihood to become nonactive depends on the pore space around it.
		
		\item[Model 3:] $k_\text{f}(\phi)$ and $A(\kappa)$. The secretory behaviour of an osteoblast depends on the pore space around it, but its likelihood to become nonactive depends on the local curvature of the bone surface that it occupies.
		
		\item[Model 4:] $k_\text{f}(\phi)$ and $A(\phi)$. Both the secretory behaviour of an osteoblast and its likelihood to become nonactive depend on the pore space around it.
	\end{description}

\end{enumerate}
The functional dependences of $k_\text{f}$ and $A$ upon the geometric variables $\kappa$ or $\phi$ in each model are determined by substituting $R = -1/\kappa$ or $R = L \sqrt{\phi/\pi}$ in the functions $k_\text{f}(R)$ and $A(R)$ determined in the first step, see Eqs~\eqref{kf-irreg},\eqref{A-irreg}. All four models result in identical behaviour in circular pore geometries, but not in irregular pore geometries. A parametric study of cell diffusivity $D$ is performed for each model.

In each model, pore infilling is assumed to stop once osteonal porosity

\begin{align*}
	\phi(t) = \frac{\text{pore area}}{L^2} = \frac{1}{2L^2}\int_0^{2\pi} R(\theta,t)^2 d\theta 
\end{align*}%
reaches the value $\phi_H = \frac{\pi R_H^2}{L^2}$, where $R_H\approx20\,\muup$m is the average Haversian canal radius, and $L \approx 300\,\muup$m is chosen large enough to fit most typical cortical resorption cavities, which have an average diameter of about~200\,$\muup$m~\citep{Martin1998,Parfitt1983}. This porosity measure is similar to the `individual osteon porosity' defined by the ratio of pore area and initial pore area~\citep{Metz2003}, but it has the advantage of being able to compare the initial porosity of different resorption cavities, and of not being 1 initially, which helps regularise mechanical estimates (see Discussion). We note here that choosing larger values of $L$ may be needed to define $\phi$ when considering abnormally large osteons. Such a rescaling of porosity does not change the behaviours of the models since this rescaling is automatically compensated for in the functions $k_\text{f}(\phi)$ and $A(\phi)$ due to the calibration of these functions to circular pore geometries (see Eqs.~\eqref{kf-irreg}-\eqref{A-irreg}). The consideration of osteonal porosity $\phi$ (i.e. the normalised pore area of a single osteon) rather than a tissue-average porosity (which would depend additionally on pore density) is appropriate to investigate the influence of pore shape on cell behaviour within single osteons.

To assess which model represents typical evolutions of irregular pores during their infilling, we define a discrepancy measure based on the circularity of the final interface shape $R_\text{end}(\theta)$ when infilling has completed. Since Haversian canals are more regular and circular than initial resorption cavities~\citep{Parfitt1983,Metz2003} we define the discrepancy
\begin{align}
	\epsilon = \frac{1}{N}\sum_{i=1}^{N} \left[ R_\text{end}^i - R_\text{H}\right]^2    \label{eqAveR}
\end{align}    
where $R_\text{end}^i = R_\text{end}(\theta_i)$ is the radius of the final interface at the angular discretisation point $\theta_i$, and $N$ is the number of discretisation points along the pore interface. Since infilling continues in the model until the target porosity $\phi_H$ is reached, $\epsilon$ measures only deviations from circularity in the final shape, and not deviations in porosity.

\paragraph{Three-dimensional vs two-dimensional parameter values}
To convert values of three-dimensional quantities to two-dimensional values in the cross-section, we use a nominative cross-section thickness of $\Delta z = 20\,\muup$m, corresponding roughly to the size of an osteoblast. For example, an osteocyte density of 31,250mm$^{-3}$ corresponds to the value 31,250 mm$^{-3} \times \Delta z= 625/\text{mm}^2$ in the cross section. Likewise, osteoblast surface density becomes $\rho \Delta z$ in the cross-section, and cell secretory rate becomes $k_\text{f}/\Delta z$. In the following we will refer to values converted to two dimensions by this procedure.

\paragraph{Numerical simulations}
Eqs~\eqref{eqn_v}--\eqref{eqn_Rho} are solved numerically using the same techniques as~\cite{Alias2017}. A straightforward finite difference upwind scheme is used at high diffusivities, but a high-resolution finite volume method (Kurganov--Tadmor scheme) is used at low diffusivities to prevent significant numerical loss of cells. We refer the reader to~\cite{Alias2017} for more detail on these numerical schemes.

The initial resorption cavity determines the initial cavity radius $R(\theta, 0) = R_0(\theta)$. The initial osteoblast density $\rho(\theta, 0) = \rho_0 \approx 161/$mm was assumed to be the same homogeneous value in all simulations, so that in circular pores of initial radius $R_0=100\,\muup\text{m}$, the initial normal velocity of the interface is $v_0 = 1.9\,\muup\text{m}/\text{day}$, consistently with experimental data (see Sect.~\ref{Section_data}).

\subsection{Experimental data}
\label{Section_data}
{\renewcommand{\arraystretch}{1.2} 
	\begin{table*}[tb!]
		\captionsetup{justification=centering} 
		\caption{Scaled data on cell density, interface velocity (MAR), and cell secretory rate $k_\text{f}$ at different radii of infilling remodelling cavities, based on measurements reported in dogs by~\cite{Marotti1976} and the procedure outlined by~\cite{Buenzli2014a} to rescale dog data onto human data. Conversions to two-dimensional values are based on an assumed cross-section thickness $\Delta z = 20\muup$m}	
		\label{table_kf}
		\centering
		\begin{tabular}{ccccc}
			\hline
			\multicolumn{1}{c}{pore radius $R$} &
			\multicolumn{2}{c}{osteoblast density $\rho$} &
			\multicolumn{2}{c}{cell secretory rate $k_\text{f}$} \\
			\hline
			& 3D & 2D & 3D & 2D \\
			$[\muup$m$]$ & $[$mm$^{-2}]$ & $[$mm$^{-1}]$ & $[$mm$^3/$day$]$ & $[$mm$^2/$day$]$\\
			\hline
			20   & 2300 & 46 & 0 & 0\\ 
			22.9 & 3600 & 72 & $112.5 \times 10^{-9}$ & $5.63 \times 10^{-6}$\\
			43.3  & 7016 & 140 &  $128.75 \times 10^{-9}$ & $6.44 \times 10^{-6}$\\
			92.7  & 8000 & 160 & $225 \times 10^{-9}$ & $11.25 \times 10^{-6}$\\
			\hline
		\end{tabular} 
	\end{table*}
}

Experimental data measuring osteon infilling dynamics comes mostly from double labelling experiments. These experiments enable the estimation of the speed of the bone interface $v$ as a function of mean radius $R$ (or mean area)~\citep{Metz2003}. There is little literature, however, on osteoblast density $\rho$ in infilling remodelling cavities of different sizes. These two types of data ($v$ and $\rho$) determine $k_\text{f}$ by Eq.~\eqref{eqn_v}. Due to the need to use different experimental methods to determine these quantities, they are not usually collected simultaneously on the same samples. Here, we gather data from experiments conducted on animal from different species, and rescale these data to typical dimensions seen in human bone samples according to known cross-species differences, as was done in ~\cite{Buenzli2014a}. 

\paragraph{Cell density and secretory rate} \cite{Marotti1976} have measured both $v(R)$ and osteoblast density $\rho(R)$ in infilling remodelling cavities of different radii $R$ in dogs, which was used to deduce $k_\text{f}(R)$ by Eq.~\eqref{eqn_v}. Following~\cite{Buenzli2014a}, we scaled dog pore radii to human values by a linear transformation. Cell secretory rate $k_\text{f}^\text{dog}$ was scaled by a factor $1.25$ to account for higher secretion rates in humans \citep{Polig1990,Buenzli2014a}, while osteoblast density was scaled by the inverse factor $1/1.25$~\citep{Buenzli2014a}. Table~\ref{table_kf} summarises the scaled experimental data on $\rho(R)$ and $k_\text{f}(R)$. The first line of data in Table~\ref{table_kf} corresponds to quiescent osteoblasts lining the bone surface after bone formation has completed~\citep{Parfitt1983}.

\begin{figure}
	\captionsetup{justification=centering}
	\centerline{
		\includegraphics[trim={0 0 0 0}, width=0.8\linewidth,clip]{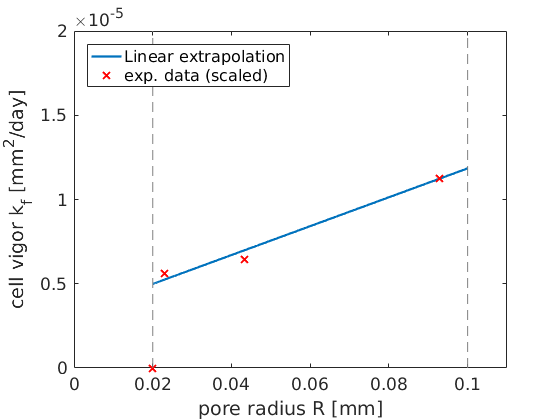}}
	
	\caption{Scaled data on (nonzero) cell secretory rate $k_\text{f}(R)$ from Table~\eqref{table_kf}, and the linear interpolation in Eq.~\eqref{eqn_kf}}
	\label{fig_kf}
\end{figure}

We interpolate the data $k_\text{f}(R)$ in Table~\ref{table_kf} linearly in $R$ between the average human Haversian canal $R_\text{H} = 20\,\muup$m (excl.) and the initial cavity radius (or cement line radius) $R_\text{c}=100\,\muup$m~\citep{Martin1998} as
\begin{align}
	k_\text{f}(R) = a_{k_\text{f}} + b_{k_\text{f}}R, \label{eqn_kf}   
\end{align}
where for $k_\text{f}(R)$ in mm$^2/$day, $a_{k_\text{f}} = 3.2741 \times 10^{-6}\,\text{mm}^2/\text{day}$ and $b_{k_\text{f}} = 8.5728 \times 10^{-5}\, \text{mm}^2/\text{day}$ (see Fig.~\ref{fig_kf}).

\paragraph{Interface velocity (matrix apposition rate)}		
Data on the velocity of the bone formation front is much more abundant. To take advantage of this abundance, we use a more extensive dataset collected on sheep by~\cite{Metz2003}, whom report the percentage of bone infilled versus cavity radius. We rescale this data onto human values by a similar linear transformation, such that a 100\% unfilled cavity corresponds to the initial cavity radius $R_c=100\muup$m, and a 0\% unfilled cavity corresponds to the Haversian canal radius $R_H=20\muup$m, as was also done in~\cite{Buenzli2014a}. This scaled data is shown in Fig.~\ref{fig_Ave} along with simulation results.

\section{Results}

\label{Section_Results}

\paragraph{Cell depletion rate in circular pore geometry}
The geometric regulation of cell depletion rate $A(R)$ in a perfectly circular infilling pore is determined by comparing the osteoblast surface densities $\rho(R)$ reached by the simulations at different radii with the data from Table \ref{table_kf}. All the numerical simulations in this circular geometry assume that cell secretory rate $k_\text{f}(R)$ is the function given in Eq.~\eqref{eqn_kf}. In the circular geometry, cell diffusion is irrelevant so long as the initial confluence of osteoblasts is achieved with a uniform density before they become active, which is assumed here. 

\begin{figure}
	\captionsetup{justification=centering}

	\centerline{
		\includegraphics[trim={0 0 0 0}, width=0.8\linewidth,clip]{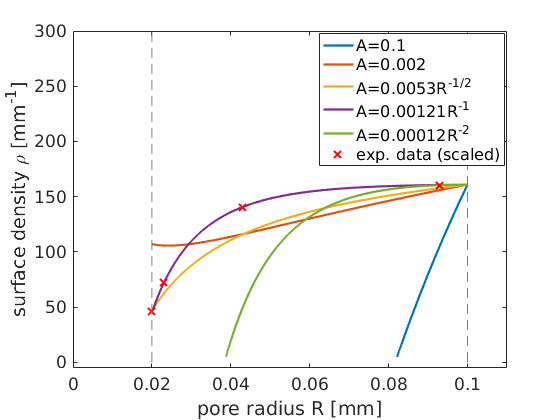}
	}

	\caption{Comparison of cell surface density between model with various cell depletion rate $A$ and the 2D-human-scaled experiment data from~\cite{Parfitt1983}, along the infilling of a idealised circular osteon produced by the model $k_\text{f}$ in Eq.~(\ref{eqn_kf})}
	\label{fig2}
\end{figure}

Figure~\ref{fig2} shows that a constant cell depletion rate does not lead to cell densities in the simulations that match the in-vivo cell density data from Table~\ref{table_kf}. The constant value $A=0.1\,\text{mm}/\text{day}$ used in our previous analysis of bioscaffold tissue growth~\citep{Alias2017} leads to a rapid depletion of active cells and incomplete bone formation. The value $A=0.002\,\text{mm}/\text{day}$ decreases density too fast initially (large $R$), but too slowly towards the end of bone formation (small $R$), where crowding of cells $\propto 1/R$ induced by the shrinking pore surface area takes over. With this value of $A$, the density of quiescent cells lining the bone surface at completion of bone formation is twice larger than measured values. 

To match the nonlinear decrease in cell surface density with decreasing cavity radius $R$ despite the strong crowding of cells that occurs at small $R$, it is necessary to increase cell depletion as $R$ decreases. Testing power-law dependences of $A$ upon $1/R$, we find that an excellent fit of the simulation to the data is obtained by choosing
\begin{align}
	A(R)= \frac{A_0}{R}   \label{eqn_A}
\end{align}
with $A_0 =  0.00121\,\text{mm/day}$ (Fig.~\ref{fig2}). Clearly, in this case, simulations also reproduce the dynamics of the interface $v(R)$ measured by~\cite{Marotti1976}, by Eq.~\eqref{eqn_v}. When compared with the independent data $v(R)$ measured by~\cite{Metz2003}, there is only a slight deviation from the average behaviour that remains within the experimental variability (see Fig.~\ref{fig_Ave}). The bone formation period required to infill the circular pore with Eq.~\eqref{eqn_A} is about 80 days, which is consistent with reported durations of 3 months mentioned in~\cite{Parfitt1994} and ~\cite{Martin1998}.

These results suggest that in regular bone pores of circular cross-sections, the geometric regulation of the individual behaviours of osteoblasts is such that as cavity radius $R$ decreases, cell secretory rate decreases linearly with $R$ by Eq.~\eqref{eqn_kf}, and cell depletion rate increases as $1/R$ by Eq.~\eqref{eqn_A}. With these individual cell behaviours, the collective crowding of cells induced by the shrinking pore surface area, and with the generation of osteocytes, osteoblast density decreases nonlinearly as the pore infills (Fig.~\ref{fig2}), while the velocity of the interface (matrix apposition rate) decreases roughly linearly with $R$ (Fig.~\ref{fig_Ave}).

\paragraph{Individual cell behaviours in non-circular pores}

Models~1--4 are proposed as possible generalisations of the behaviours of $k_\text{f}$ and $A$ in noncircular geometries (see Section~\ref{Section_model}). These models are obtained by interpreting $R$ either by means of curvature~$\kappa$ or porosity~$\phi$ in Eqs~\eqref{eqn_kf} and~\eqref{eqn_A}. Doing so results in the following possible geometric regulations of cell secretory rate and cell depletion rate:
\begin{alignat}{2}
	&k_\text{f}(\phi) = a_{k_\text{f}} + \frac{b_{k_\text{f}}L}{\sqrt{\pi}}\sqrt{\phi},
	\quad&& k_\text{f}(\kappa) = \begin{cases}a_{k_\text{f}} - \frac{b_{k_\text{f}}}{\kappa}, &\text{if } \kappa \leq \kappa_c < 0\rule[-1.5ex]{0pt}{1.5ex}
		\\a_{k_\text{f}} - \frac{b_{k_\text{f}}}{\kappa_c}, &\text{if } \kappa_c < \kappa < 0\rule[-1.5ex]{0pt}{1.5ex}
		\\0, &\text{if } 0 \leq \kappa 
	\end{cases}\label{kf-irreg}
	\\
	& A(\phi) = \frac{A_0 \sqrt{\pi}}{L\sqrt{\phi}},
	&& A(\kappa) = \begin{cases}-A_0\kappa, &\text{if }\kappa < 0
		\\0, &\text{if } 0 \leq \kappa
	\end{cases}\label{A-irreg}
\end{alignat}
where $\kappa_c = -1/R_c$ is the curvature of the cement line in the circular geometry. These expressions all recover Eqs~\eqref{eqn_kf} and~\eqref{eqn_A} when $\kappa = -1/R$ and $\phi = \pi R^2/L^2$, so long as $R\leq R_c$. Cell secretory rate has been bounded from above when controlled by curvature on portions of the bone substrate where $\kappa_c<\kappa<0$, i.e., on portions that are flatter than the cement line in circular geometry, due to the limited capacity of cells to secrete new bone matrix. In convex regions of the bone substrate ($\kappa \geq 0$), both $k_\text{f}(\kappa)$ and $A(\kappa)$ are extrapolated to be zero. It is assumed that when curvature controls these behaviours, osteoblasts in these regions become quiescent cells with $k_\text{f}=0$ and $A=0$, as suggested by bioscaffold tissue growth experiments~\citep{Rumpler2008,Bidan2012,Bidan2013}. The meaning of Eqs. \eqref{kf-irreg},\eqref{A-irreg} in regard to Model 1--4 is described in page 4.  

The extrapolation of the geometric regulations of individual cell behaviours to noncircular pore shapes in Eqs~\eqref{kf-irreg} and~\eqref{A-irreg} now entirely defines Models 1--4 with the evolution equations Eqs~\eqref{eqn_v}--\eqref{eqn_Rho}. The only free parameter that remains in the four models is the cell diffusivity $D$.

\begin{figure}
	\captionsetup{justification=centering}
	
	\centerline{
		\includegraphics[trim={0 0 0 0},width=0.8 \linewidth,clip]{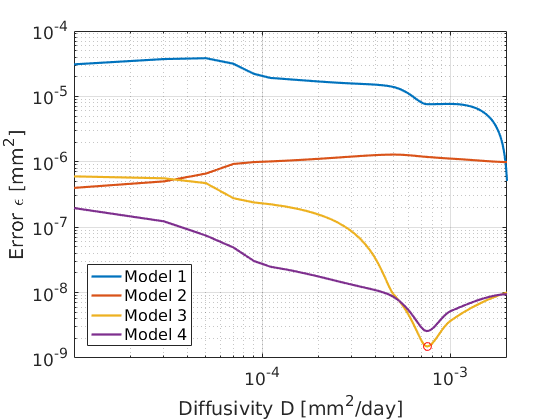}
	}
	
	\caption{Errors calculated for Models 1, 2, 3 and 4 with different diffusivity $D$, plotted using the log-log scale. Minimum error is obtained when $D \approx 0.00075$ and when using Model 3}
	\label{figerror}
\end{figure}

\begin{figure*}
	\captionsetup{justification=centering}

	\centerline{
		\includegraphics[trim={30 100 30 0} ,width=0.6\linewidth,clip]{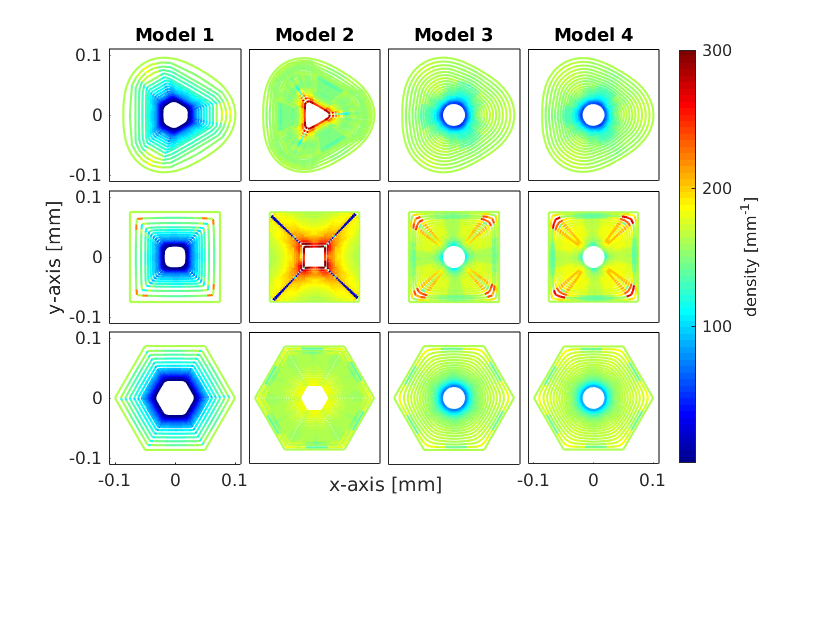}}
	
	\caption{Infilling of various initial pore shapes (cosine, square, and hexagonal) simulated by Models~1--4 with $D=0.00075\,\text{mm}^2/\text{day}$. The pore interface is colored according to cell density, shown at regular time intervals}
	\label{fig_smoothinterface}
\end{figure*}

\paragraph{Influence of cell diffusivity}
In a circular pore with homogeneous initial osteoblast density, cell diffusion has no effect on the evolution, so that cell diffusivity $D$ cannot be estimated. Our previous simulations of tissue growth in bioscaffold pores exhibited strong qualitative changes in growth patterns driven by diffusivity~\citep{Alias2017}. To assess the influence of cell diffusivity on the infilling of bone pores in noncircular geometries, we perform a parametric study of $D$ on pores of square, hexagonal, and cosine shapes (the cosine initial pore shape is similar to a smoothed triangle) of identical perimeter 0.6\,mm, so that there is the same number of osteoblasts at the onset of bone formation. Tissue growth within such pore shapes has been investigated in bioscaffold experiments. While these regular pore shapes are not realistic bone remodelling cavities, they enable us to understand the influence of the sharpness of a cusp in the initial interface without the confounding influence of other irregularities.

The measures of discrepancy from circularity calculated for each of these shapes at the end of bone formation by Eq.~\eqref{eqAveR} are summed and plotted in Fig.~\ref{figerror} as a function of $D$ for each model. The minimum total discrepancy $\epsilon$ occurs for Model~3 when $D \approx 0.00075$\,mm$^2$/day. However, the total discrepancies $\epsilon$ in Model~3 and Model~4 are very similar and not significantly different from the minimum when $D \gtrsim 0.0003\,\text{mm}^2/\text{day}$. 

\begin{figure}
	\captionsetup{justification=centering}
	
	\centerline{
		\includegraphics[trim={35 50 60 50} ,width=1\linewidth,clip]{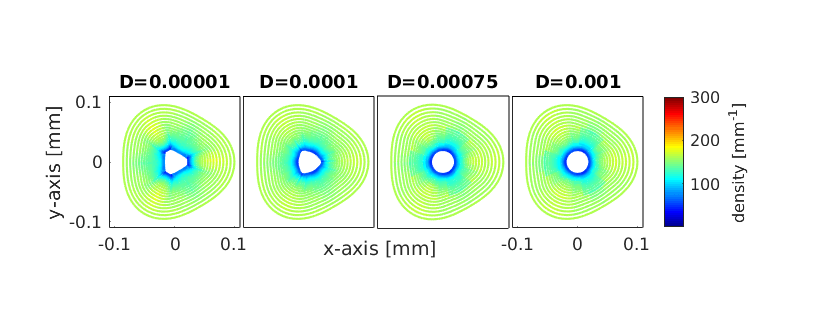}}
	
	\caption{Influence of diffusivity $D$ (in $\text{mm}^2/\text{day}$) on the infilling of the cosine pore simulated by Model~3}
	\label{fig_diffusion}
\end{figure}

Figure~\ref{fig_smoothinterface} shows the evolution of the interface in each Model at the fixed value of diffusivity $D=0.00075\,$mm$^2$/day. It is clear from Figs~\ref{figerror} and~\ref{fig_smoothinterface} that Models~1 and~2, in which cell secretory rate is assumed to be controlled by curvature, do not lead to an efficient smoothing of the interface. In contrast, the final interfaces for Models~3 and~4 are all roughly circular and indifferentiable regardless of the initial corner angles. Note that Models~3 and~4 do not smooth out initial corners efficiently if diffusion is too low (Fig.~\ref{figerror}). The influence of diffusivity on the evolution of the cosine pore interface is shown for Model~3 in Fig.~\ref{fig_diffusion}.

\paragraph{Application to real osteonal geometry}
New bone formed during the infilling of cortical pores is lamellar. In this type of bone, past locations of the bone interface are recorded as lamellae boundaries visible experimentally in histological slices. These boundaries provide a point of comparison with simulations of our mathematical model. 

\begin{figure*}[t!]
	
	\captionsetup{justification=centering}
	\centering\includegraphics{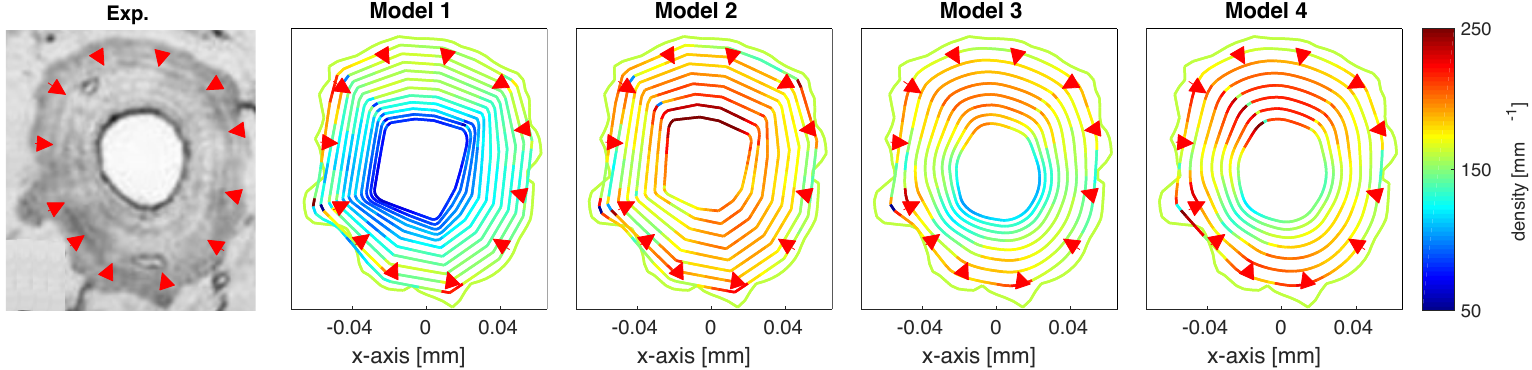}
	\caption{Image of a real osteon seen in a histological cross section of human cortical bone (reproduced with permission from \cite{Skedros2005}) and corresponding simulations of pore infilling using Models~1--4. Arrowheads indicate the boundary between two lamellae in the histological image. Simulated interfaces are coloured according to osteoblast density and shown every 4.56 days}
	\label{fig_compare_osteon}
\end{figure*}

Figure~\ref{fig_compare_osteon} compares simulations of cortical pore infilling starting from the cement line of a real osteon~\citep{Skedros2005}. The initial pore cavity assumed in the simulations is the boundary between the light grey region at the edges of the histological image (old bone) and the darker grey region (lamellar bone formed by infilling). Lamellae boundaries in the new bone are seen as faint, thin dark lines. We have indicated the approximate boundary between two lamellae with arrowheads at about one third of the new bone wall thickness. The simulations shown in Fig.~\ref{fig_compare_osteon} were all performed with a cell diffusivity $D=0.00075$ mm$^2$/day, a uniform initial surface density of osteoblasts $\rho_0 = 161/\text{mm}$, and a uniform osteocyte density $\text{Ot}_f = 625/\text{mm}^2$. The simulations were stopped once the infilling pore reached the same porosity as that of the histological image in Fig.~\ref{fig_compare_osteon}.

Pore interfaces simulated with Models~3 and~4 match the experimental lamellae boundary (arrowheads) very well, despite the irregularity of the initial interface and the lack of experimental information on the initial density of osteoblasts in the histological image. As time proceeds, the divergence between simulated interfaces and real lamellar boundaries increases. The final pore shape obtained by Models~3 and~4 is regular, but has some difference to the final pore shape in the experimental image. This can be expected from a dynamic system's perspective as initial errors are likely to amplify without regulatory mechanisms. There is little qualitative difference between Model~3 and Model~4. Model~3 leads to a slightly more homogeneous osteoblast density lining the final pore interface.

\paragraph{Comparison with double labelling data}
To understand more thoroughly how efficiently variations in the initial pore interface are smoothed, and how these variations affect the speed of new bone formation, we generated 20 virtual initial pore interfaces by randomly perturbing the radius of the interface between the values $R_\text{min}=0.06$ mm and $R_\text{max}=0.12$ mm according to $R_0(\theta) = R_\text{min} + \overline{\zeta(\theta)}(R_\text{max}-R_\text{min})$ with uniformly distributed random noise $\zeta(\theta)\in (0,1)$ smoothed by Matlab's \texttt{loess} method using local regression based on the weighted linear least squares and a polynomial model to provide $\overline{\zeta(\theta)}$~\citep{Matlab2016b}.

\begin{figure*}
	\captionsetup{justification=centering}
	
	\centering
	\includegraphics[trim={0 110 85 0} ,width=0.8\linewidth,clip]{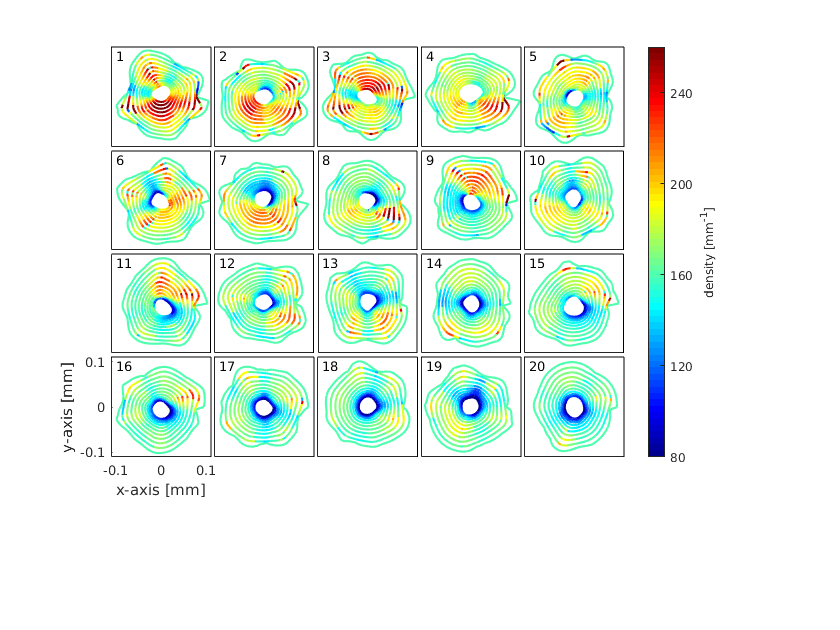}
	\caption{Infilling of random irregular pores resembling osteons using Model 3 and $D=0.00075$. The pore interfaces are shown at regular time intervals and coloured according to cell density. The pores are organised by the time it requires to infill them to a porosity of 5\%, and numbered accordingly from~1 (fastest infilling) to~20 (slowest infilling), see also Fig.~\ref{fig_Ave}}
	\label{fig_three_osteons}
\end{figure*}

\begin{figure}[t]
	\captionsetup{justification=centering}
	
	\centerline{
		\includegraphics[trim={0 0 0 0},width=1.05 \linewidth,clip]{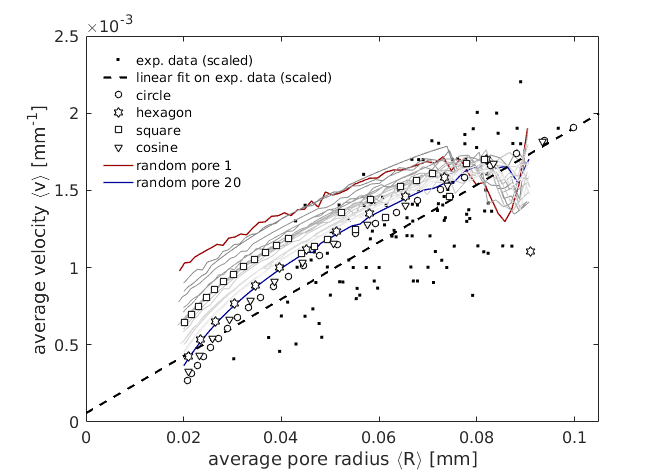}
	}
	
	\caption{Comparison between experimental double labelling data on matrix apposition rate (black squares) and simulation data. Experimental data are based on measurements from~\cite{Metz2003}, appropriately scaled to match human resorption cavity dimensions, see~\cite{Buenzli2014a}. Simulation data are calculated as the average interface velocity versus the average pore radius during simulation runs of Model~3 with $D=0.00075\,\text{mm}^2/\text{day}$ starting from a variety of initial pore shapes: square pore (open squares), hexagonal pore (open stars), cosine pore (open triangles), circular pore (open circles), and the~20 random pores of Fig.~\ref{fig_three_osteons} (solid red line for random pore 1, solid blue line for random pore 20, and solid grey lines for random pores 2--19)}
	\label{fig_Ave}
\end{figure}

The purpose of this population of initial pore shapes is to help understand the experimental variability seen in double labelling data~\citep{Marder2011}. Figure~\ref{fig_three_osteons} shows simulations of the infilling of these random pore shapes using Model~3 with $D=0.00075\,\text{mm}^2/\text{day}$. The final interfaces are all roughly circular with little inhomogeneity in cell surface density despite the varied initial pore shapes. The pores are organised and numbered according to how fast they refill.

In Figure~\ref{fig_Ave}, the average velocity of the interface in these simulations is shown versus average pore radius, and compared with double labelling experimental data. The average instantaneous velocity is estimated in the simulations as
\begin{align}
	\overline{v}(t) = \frac{|A'(t)|}{P(t)} \approx \frac{|\Delta A(t)|}{\Delta t P(t)} 
	\label{eqAveV},
\end{align}
where $A(t)$ is the pore area, $\Delta A$ is the change in pore area during the time increment $\Delta t$, and $P(t)$ is the pore perimeter. Simulation results are in good agreement with the linear regression line of the experimental data, particularly as the model has not been fitted to this experimental data. The average velocity of the interface is spread over a range of values around the regression line at an early stage of bone formation (large average radius). Irregular initial pores with highly curved interfaces tend to fill in quicker overall than more circular pores, as can be seen by identifying the highest and lowest average velocity curves with the initial shapes in Fig.~\ref{fig_three_osteons}.

This observation is corroborated by the infilling simulations of the regular pore shapes. The square initial pore (cyan dots in Fig.~\ref{fig_Ave}), which has the sharpest corners, infills the fastest, followed by the hexagonal, cosine, and circular pore shapes. In the square and hexagonal pore shapes, there are large sections of zero curvature where osteoblasts secrete new bone, but are not depleted in Model~3. However, when the infilling of these pore shapes is simulated with Model~4, where cell depletion rate occurs uninterruptedly as it depends on porosity rather than curvature, there is little difference in the average velocity curves (data not shown), which means that the mode of geometric control of cell depletion plays a subdominant role for the infilling rate compared to cell crowding.

\section{Discussion}
\label{Section_Discussion_Conclusions}

Bone remodelling is regulated at many scales by a variety of mechanisms of different nature, including biochemical, mechanical, and geometrical~\citep{Parfitt1983,Martin1998,Sims2014,Lerebours2016a}. At the tissue scale, the availability of bone surface area is an important factor that influences the propensity of bone renewal, and in particular, the rate and location of bone loss in  osteoporosis~\citep{Martin1972,Martin1984,Buenzli2013,Pivonka2013,Lerebours2016a}. In this work, we have investigated the geometric regulation of bone remodelling cavities at a lower scale, the scale of cell--tissue interaction, using a comprehensive population model of osteoblasts and experimental data on cortical bone formation dynamics.

The geometric regulation of tissue-synthetising cells at the cell--tissue scale has been exhibited in many in-vitro experiments~\citep{Rumpler2008,Bidan2012,Bidan2013,Guyot2014,Guyot2015,Guyot2016,Nelson2005,Dunlop2010,Bidan2013b,Gamsjager2013,Knychala2013}, but it remains difficult to understand the precise mechanisms by which geometry constrains tissue growth patterns. One difficulty is to disentangle the influence of geometry on the collective behaviour of cells and the influence of geometry on the individual behaviour of cells. Another difficulty is to determine what geometric variables are influencing cell behaviour, particularly as geometric features such as curvature and porosity involve length scales that are much larger than individual cell bodies. In this paper, we have addressed these two difficulties by a mathematical modelling approach.

A direct control of curvature onto single osteoblasts (e.g., via focal adhesions) may occur at the onset of new bone formation as osteoblasts may line a rough bone surface made of Howship's lacunae eroded by bone-resorbing cells~\citep{Martin1998}. However, soon after Howship's lacunae are filled and smoothed, typical radii of curvature of infilling pores in cortical bone range from $R_c\approx 100\,\muup\text{m}$ at the start of bone formation to $R_H\approx 20\,\muup\text{m}$ at the end of bone formation, while osteoblasts have an approximate size of about $20\,\muup\text{m}$. In~\cite{Alias2017}, we have proposed that tissue-forming cells are still able to sense such large geometrical features of the tissue substrate dynamically, by the collective crowding or spreading influence of curvature onto cell density. Even at the end of formation, Haversian canal perimeters of $\sim 250\,\muup\text{m}$ are able to hold about 12~osteoblasts, so that small differences in osteoblast density around the perimeter are possible. Histologically, there may be irregularities in the canal's interface at lower scales, that are covered up by osteoblasts. These lower-scale irregularities may also affect osteoblasts, but there is little quantitative data that allow comparison with mathematical models. The influence of interface curvature in our model only refers to pore shape irregularities at length scales higher than about $20\muup$m. Other curvature-dependent mechanisms than dynamic changes in density have been proposed, such as the tissue surface tension of actin networks~\citep{Nelson2005,Rumpler2008,Bidan2012}. However, due to the fast primary mineralisation of bone~\citep{Parfitt1983,Martin1998}, it is unlikely that bone tissue surface tension plays a significant role during bone formation.

Interestingly, our simulations suggest that the pore infilling dynamics of Models~1 and~2, in which cell secretory rate $k_\text{f}$ is influenced by curvature, is not smoothing irregularities of the interface very well (Figs~\ref{figerror},~\ref{fig_smoothinterface}). Cortical pore infilling results in Haversian canals that are much smoother and regular than initial cement lines~\citep{Parfitt1983}. Our simulations thus suggest that cell secretory rate may be controlled not by curvature, but by the porosity of the infilling cavity as assumed in Models~3 and~4.

A porosity control of the individual behaviour of osteoblasts is harder to conceptualise than a curvature control, since it cannot be ascribed to osteoblasts sensing local density changes~\citep{Alias2017}. Curvature exerts a direct influence on local cell density changes, but not porosity (see Eq.~\eqref{eqn_Rho}). The ability of osteoblasts to perceive porosity changes indicates a larger scale of intercellular signalling. It is well-known that bone formation is regulated mechanically by the network of osteocytes embedded within bone~\citep{Cowin1998,Turner1994,Burger1999,Sikavitsas2001,Bonewald2011,Buenzli2015b,Lerebours2016b}. This network of cells is in direct contact with the layer of osteoblasts lining the bone surface~\citep{Marotti2000,FranzOdendaal2006}. It is also known that microscopic mechanical strains of bone matrix are determined mostly by porosity~\citep{Grimal2011,Granke2012,Rohrbach2012}. A simple micromechanical model of stress concentration~\citep{Lerebours2016b} shows indeed that the strain energy density of bone matrix $\Psi$ is given by
\begin{align*}
	\Psi = \frac{\frac{1}{2}\mathbb{C}^{\text{micro}^{-1}}_\text{bm} \left(F/L^2 \right)^2}{(1-\phi)^2},  
\end{align*}
where $\mathbb{C}^{\text{micro}^{-1}}_\text{bm}\approx 0.0482 \ \text{GPa}^{-1}$ is the inverse longitudinal bone matrix stiffness, and $F/L^2$ is the compressive stress exerting onto the osteonal region $L^2$. Our finding that cell secretory rate may depend on porosity rather than curvature may therefore indicate that during pore infilling, osteocytes respond to decreasing local mechanical strains by sending inhibitory signals to osteoblasts. This may occur e.g.\ via sclerostin inhibition of the Wnt pathway~\citep{Marotti2000,Martin2000,Bonewald2008,Buenzli2012}. This mechanics-induced inhibition is consistent with the mechanical control of bone adaptation~\citep{Lerebours2016b}, and may also act as a stopping mechanism for bone formation when local mechanical strains fall below a setpoint threshold.

Other porosity-dependent mechanisms than mechanical strains of bone matrix are possible. For example, fluid flow within Haversian pores has been shown to trigger bone formation and prohibit bone loss \citep{Cowin2015,Qin2003}, even without the presence of osteocytes \citep{Kwon2010,Kwon2012}. \cite{Martin1998} have suggested that the decreasing space between osteoblasts and the blood vessel running within cortical pores might also signal osteoblasts to slow down, and perhaps stop, bone deposition during osteon infilling. However, if so, one would expect to find differences in the activity of osteoblasts around the pore's perimeter in irregular pore shapes. Our finding that cell secretory rate depends on porosity thus excludes this model, because porosity has a uniform value in the cross section.

Our simulations do not enable us to clearly disentangle the nature of geometric regulation of cell depletion. Model~3 and Model~4, which assume curvature-dependent and porosity-dependent cell depletion rate respectively, result in similar pore infilling dynamics (Figs~\ref{figerror}, \ref{fig_smoothinterface}, \ref{fig_compare_osteon}). Both models assume that the density of osteocytes generated at the bone deposition front is uniform. While radial dependences of osteocyte density within osteons have been reported~\citep{Power2010,Hannah2010}, it is unclear if there are also angular osteocyte density inhomogeneities in irregular osteons. High-resolution microCT scans of bone samples could be used to investigate such inhomogeneities~\citep{Hannah2010,Dong2014}. A dependence upon interface curvature of osteocyte density (e.g., induced by an implicit curvature dependence of cell burial rate~\citep{Buenzli2015}) would make the differentiation of osteoblasts into osteocytes curvature-dependent too in Eq.~\eqref{eqn_Rho}. This would affect the geometric regulation of cell depletion rate $A$ determined in Fig.~\ref{fig2} and could thereby result in more pronounced differences between Model~3 and Model~4.

With a porosity control of cell secretory rate, one may understand the variability of double labelling data on average interface velocity (MAR) as being due to the level of irregularity of the initial resorption cavities and the strong influence of cell crowding in highly curved concavities of the bone interface. This is shown by our simulations where the fastest infilling pores shapes in Fig.~\ref{fig_Ave} are those in which small concavities of the initial interface concentrate cells strongly (e.g., random shapes no.~1 and~2 in Fig.~\ref{fig_three_osteons}). Conversely, the slowest infilling pore shapes in Fig.~\ref{fig_Ave} are those that have a more circular initial interface (e.g., random shapes no.~19 and~20 in Fig.~\ref{fig_three_osteons}). All these random shapes have the same initial porosity. These results may be put in perspective with experimental studies showing that osteon circularity correlates positively with age and negatively with strain in humans \citep{Britz2009,Hennig2015} and animals \citep{Skedros2004}. While it is unclear whether the age-related changes can be linked to changes in mechanical strains due to reduced physical activity~\cite{Hennig2015}, our study suggests that osteons may infill slower with age due to being more circular, and faster in mechanically stimulated bone due to being less circular.

Despite porosity controlling secretory rate in Models~3 and~4, the overall porosity of the initial pore influences the average infilling rate less than the presence or absence of highly curved concavities. Figure \ref{figS1} in the appendix shows simulations of random pore $8$ scaled down by $70\%$ and scaled up by $130\%$. While the smaller pore infills slower than the larger pore due to the porosity dependence of the secretory rate $k_\text{f}$, the difference in average interface velocity (MAR) (Fig. \ref{figS2} in the appendix) is less than that induced by differences in the irregularity of the initial pores (Figs. \ref{fig_three_osteons}-\ref{fig_Ave}). The nonlinearity of the curvature-induced cell crowding makes this influence dominant for the overall speed of pore infilling.

Cortical pore infilling is a complex biological process and some bone histomorphometric studies have shown that bone formation sometimes occurs with a different pattern than the usual regular infilling that we have assumed in this paper. For example, bone formation in some osteons may pause during refilling~\citep{Martin1998}. Other osteons may not infill, but drift sideways 
~\citep{Robling1999,Maggiano2011,Crowder2011}. The lamellar structure of cortical bone seen in a cross-section may also be spiraling around the Haversian canal, or may not form a closed ring even on concave portions of the bone substrate~\citep{Pazzaglia2012}. These behaviours are not possible within our mathematical model. They are likely to require further regulatory mechanisms of active osteoblasts, such as local, inhomogeneous mechanical clues.

In summary, we have proposed a general mathematical model of pore infilling in cortical bone to investigate the geometric regulation of osteoblasts during bone formation. The novelties of this mathematical model are (i) to factor out the collective influence of geometry on crowding and spreading of bone-forming cells in order to determine the influence of geometry on invididual cell behaviours; and (ii) to use a population of initial pore shapes to understand variability in double labelling data. This approach to investigate cell behaviour and biological variability is a promising way to circumvent limitations of biological experiments. Our findings suggest that cell secretory rate is not regulated by the curvature of the bone surface, but by the porosity of the infilling cavity, for example by means of a mechanical response of the osteocytes generated during pore infilling. We also find that cell depletion rate is strongly regulated by the geometry of the infilling pore, but our model is unable to distinguish which geometrical variable is responsible for this regulation. Finally, our simulations suggest that the circularity of a pore is an influential variable for the overall speed of pore infilling, with less circular pores infilling faster than more circular pores.

\subsubsection*{Acknowledgements}

\small
We thank Prof Matthew Simpson and Prof Kevin Burrage for fruitful discussions, and the three anonymous reviewers for their suggestions. MAA is a recipient of the fellowship scheme from the National University of Malaysia, and the departmental scholarship from the School of Mathematical Sciences, Monash University, Australia. PRB gratefully acknowledges the Australian Research Council for Discovery Early Career Research Fellowship (project No.~DE130101191).
\normalsize

\begin{appendix}
	
	\appendix

	\section*{Appendix: Influence of initial porosity on average infilling rate}

	\label{appx:num}

\begin{figure}[h]
		\captionsetup{justification=centering}
		
		\centerline{
			\includegraphics[trim={35 20 55 50} ,width=1.0\linewidth,clip]{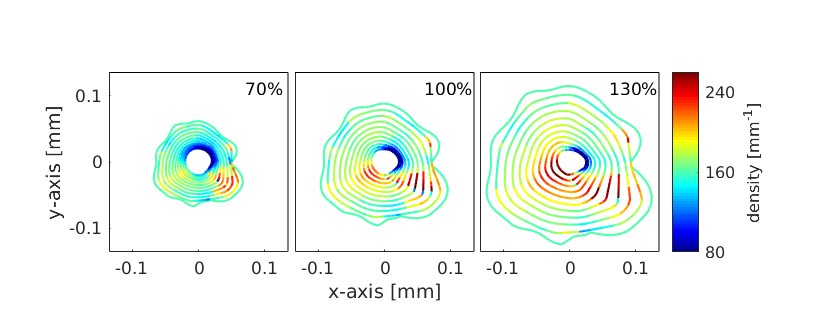}}
		
		\caption{Infilling of random pore $8$ using Model $3$ and $D = 0.00075$ at $70\%$, $100\%$ and $130\%$ of the size shown in Fig. \ref{fig_three_osteons}}
		\label{figS1}
	\end{figure}

\begin{figure}[h]
		\captionsetup{justification=centering}
		
		\centerline{
			\includegraphics[trim={0 0 0 0} ,width=1.0\linewidth,clip]{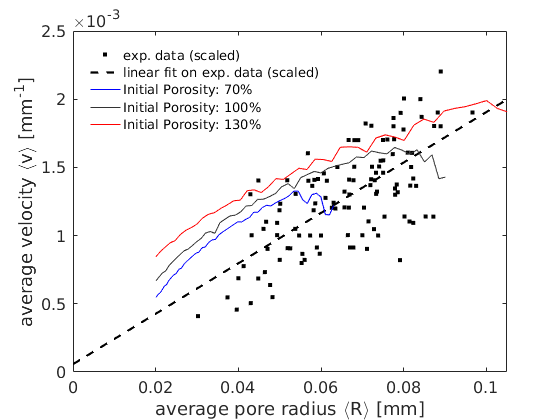}}
		
		\caption{Comparison of the average interface velocity versus the average pore radius between different scales of random pore 8 (solid lines) and experimental data (black squares)}
		\label{figS2}
	\end{figure}

	Figure \ref{figS1} shows simulations of the infilling of random pore 8 of Fig. 7 scaled down by $70\%$ and scaled up by $130\%$, performed with Model 3 and $D = 0.00075 \text{mm}^2 /\text{day}$. The corresponding plot of the average interface velocity versus the average pore radius are shown in Fig. \ref{figS2}.

\end{appendix}

%
%
%
%
%
%
%
%
%
%
%
%
%
%
%
%
%

\bibliographystyle{ieeetr}
\bibliography{reference}

\end{document}